# Few-cycle lightwave-driven currents in a semiconductor at high repetition rate

Fabian Langer,[1,2] Yen-Po Liu,[1,2] Zhe Ren,[1,2] Vidar Flodgren,[1,2] Chen Guo,[1] Jan Vogelsang,[1,2] Sara Mikaelsson,[1] Ivan Sytcevich,[1] Jan Ahrens,[3] Anne L'Huillier,[1] Cord L. Arnold,[1] and Anders Mikkelsen[1,2,*]

[1]*Department of Physics, Lund University, Box 118, 22100 Lund, Sweden*
[2]*Nano Lund, Lund University, Box 118, 22100 Lund, Sweden*
[3]*TEM Messtechnik GmbH, Großer Hillen 38, 30559 Hannover, Germany*
*Corresponding author: anders.mikkelsen@sljus.lu.se

**When an intense, few-cycle light pulse impinges on a dielectric or semiconductor material, the electric field will interact nonlinearly with the solid, driving a coherent current. An asymmetry of the ultrashort, carrier-envelope-phase-stable waveform results in a net transfer of charge, which can be measured by macroscopic electric contact leads. This effect has been pioneered with extremely short, single-cycle laser pulses at low repetition rate, thus limiting the applicability of its potential for ultrafast electronics. We investigate lightwave-driven currents in gallium nitride using few-cycle laser pulses of nearly twice the duration and at a repetition rate two orders of magnitude higher than in previous work. We successfully simulate our experimental data with a theoretical model based on interfering multiphoton transitions, using the exact laser pulse shape retrieved from dispersion-scan measurements. Substantially increasing the repetition rate and relaxing the constraint on the pulse duration marks an important step forward towards applications of lightwave-driven electronics.**

**Introduction**

Steering the quantum motion of electrons with the electric field of light has led to some of the most fascinating phenomena in photonics. For example, the interaction between intense light and gases resulted in the emission of high-order harmonic radiation [1,2] and attosecond pulses [3,4]. Accelerating electrons in solids with lightwaves enabled the observation of ballistic currents [5], high-order harmonic generation [6-8] and quasiparticle collisions [9,10]. The utilization of sub-cycle coherent currents opens the possibility to realize ultrafast circuitry driven by the electric field of light. The time scale for carrying out electronic operations is thus reduced to a half-cycle of an electromagnetic wave, possibly elevating electronics towards petahertz ($10^{15}$ Hz) clock rates.

The direct observation of macroscopic currents driven by the carrier wave of light was first reported in 2012 by Schiffrin and co-workers [11]. A residual current signal was induced by illuminating a fused silica sample with intense 3.8-fs pulses from a post-compressed 3-kHz titanium-sapphire amplifier and measured with gold electrodes patterned onto the sample. A few femtocoulomb ($10^{-15}$ C) of charge were displaced per laser shot and picked up by the external electronic circuit. The magnitude and direction of this current is sensitive to the carrier-envelope phase (CEP), which is the relative phase of the carrier wave with respect to its envelope function. An oscillating charge signal was observed by varying the CEP. This experiment sparked the idea to build electronic circuitry from dielectrics, which could be reversibly switched to a conductive state within an optical half-cycle. However, the extremely short pulse duration and low repetition rate limits the applicability and wide spread study of this concept.

Lightwave-driven currents were investigated in several material systems, for example $SiO_2$ [11], $CaF_2$ [12], $Al_2O_3$ [13], and GaN [14]. In this context, gallium nitride (GaN) is particularly interesting as it is the commercial backbone of white light LEDs and power electronics, for example. As a result, a vast knowledge is available on manufacturing devices and III-N material heterostructures with nanoscale precision [15-17]. This is highly advantageous when considering future construction of electronic devices in which well-defined barriers and connections will be important for functionality.

To observe lightwave-driven currents, the driving laser pulse has to be short enough so that the symmetry between positive and negative half-cycles can be broken by means of tuning the CEP. Several models explaining the underlying physical mechanism behind this phenomenon have been proposed [18-21]. Similarities to coherent control experiments of interfering multiphoton transitions driven in semiconductors by two-color fields [22-30] have also been pointed out.

Here, we investigate lightwave-driven currents in gallium nitride in a new parameter regime. Using 6.4-fs laser pulses at high repetition rate (200 kHz), we observe strong charge oscillations as a function of CEP. We study the scaling of the charge signal with pulse duration and electric field strength. Our results are well reproduced by a model developed by Khurgin [20], which is based on interfering multiphoton transitions in the nonlinear medium, highlighting the similarities to coherent control experiments [22-30]. Accurate modelling of the experimental data requires careful characterization of the laser pulse used in the measurements, which is provided by the dispersion-scan technique [31,32].

**Methods**

Our samples are made from commercially available, undoped GaN films that have been grown on a sapphire ($Al_2O_3$) substrate (MSE Supplies, $n < 5 \cdot 10^{17}$ cm$^{-3}$). A combination of titanium (adhesion layer, thickness 10 nm) and gold (thickness 30 nm) is evaporated onto the GaN wafer. Defined metal contacts are patterned by ultraviolet photolithography and subsequently developed. The contacts are protected against the environment and high laser powers with a thin layer (~15 nm) of sapphire deposited by atomic layer deposition. The sample design is shown in Fig. 1a (microscope image) and schematically in Fig. 1b. Typical gap sizes in the center amount to ~5 µm.

The experimental setup used to drive and observe lightwave currents is illustrated in Fig. 1c. Our laser system [33] is based on an ultrabroadband titanium-sapphire oscillator capable of delivering sub-6-fs pulses. The oscillator pulses are amplified in an optical parametric chirped pulse amplifier employing beta barium borate (BBO) crystals. The pump pulses are obtained from frequency doubling of the output of an ytterbium-doped rod-type fiber amplifier seeded by the same oscillator, thus achieving inherent optical synchronization in the parametric amplification stages. Additionally, we make use of two beam stabilization systems (TEM Messtechnik) for increased stability. A pulse shaper is used to finely tune higher-order dispersion. The amplified laser pulses pass through a pair of glass wedges (Fig. 1c), so that their CEP and dispersion can be controlled. A set of polarizers is used to adjust the peak intensity of the laser pulses. A parabolic mirror focuses the beam to a spot size of roughly 13 µm (full width at half maximum, FWHM, diameter). In the focus, we place either a BBO crystal for pulse characterization or a sample for driving electric currents. Metal contacts on the material are connected via aluminum wires (Fig. 1a) to the external measurement circuit. We align our GaN sample to the focus of the few-cycle laser pulses (Fig. 1a, red circle) using a 3-D micrometer stage with differential adjusters. Small electric currents can be measured by employing a sensitive pre-amplifier (Femto DLPCA-200) with a gain of $10^5$, that supports the full bandwidth of the high repetition rate. The pre-amplifier simultaneously converts the current signal into a voltage, which is measured by a lock-in amplifier.

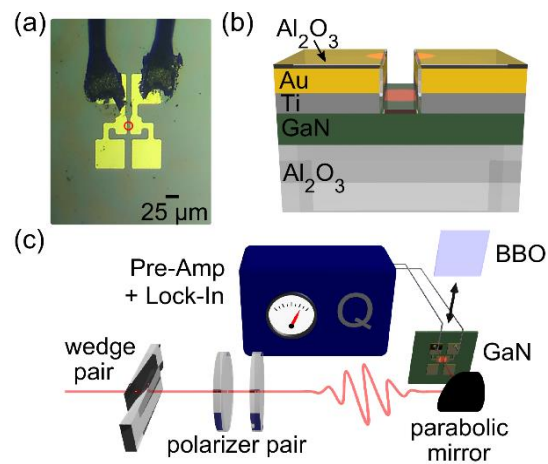

**Fig. 1. Device geometry and experimental set-up.** (a) Optical microscope image of the wire-bonded gold contacts on GaN. The laser spot position is marked by a red circle. (b) Sketch of the sample structure. Epitaxially grown GaN resides on a sapphire ($Al_2O_3$) wafer. Metal contacts are formed of a titanium adhesion layer and gold deposited directly on the GaN. A sapphire capping layer produced by atomic layer deposition protects the metal contacts. (c) Depiction of the set-up. The dispersion and power of ultrashort light pulses are controlled by a wedge and a polarizer pair, respectively. Focusing is achieved by a parabolic mirror. Either the GaN sample with metal electrodes for lightwave-driven currents or a BBO crystal for d-scan can be placed in the focus.

**Results and Discussion**

Upon illumination of the sample, we observe an oscillating current as a function of the insertion of the glass wedges, on top of a large, alignment-dependent, offset signal (Fig. 2a, red curve). While the offset signal roughly reflects the peak intensity when changing the glass insertion, the oscillating part results from the CEP change introduced by the dispersive glass. Fourier filtering is used to remove the intensity-dependent offset. The blue curve in Fig. 2a shows a clear charge oscillation as a function of CEP. The amplitude of the oscillations decreases with glass insertion as the laser pulses become too long to cause a measurable effect.

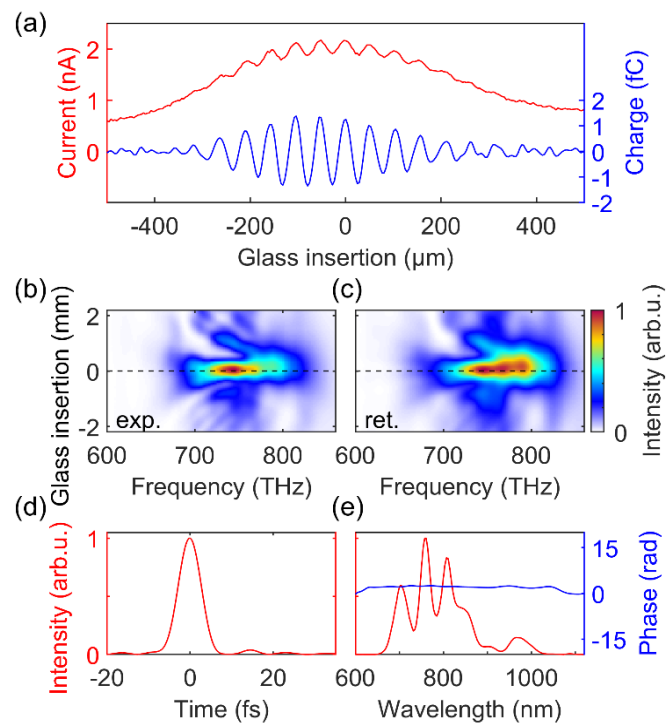

**Fig. 2. Lightwave-driven currents and d-scan.** (a) Measured and averaged current signal (red) at a pre-amplifier gain of $10^5$ when the 6.4-fs pulses are focused onto the metal gap. The blue curve shows the transferred charge per laser shot, where the offset signal has been removed by Fourier filtering. (b) Measured d-scan trace. The graph shows the spectrally resolved second-harmonic intensity as a function of glass insertion of a BK-7 wedge pair. (c) Retrieved d-scan trace. (d) Calculated temporal intensity profile of the fundamental pulses as retrieved from the d-scan measurement, revealing a FWHM pulse duration of 6.4 fs (2.4 cycles). (e) Retrieved intensity spectrum with phase.

Since the transferred charge depends sensitively on the exact shape of the driving pulses, a careful characterization is crucial. Thus, the dispersion-scan (d-scan) is recorded at the same position where the measurement is performed (Fig. 1c). Fig. 2b and c show examples of measured and retrieved d-scan traces. The extracted pulse profile and intensity spectrum can be seen in Fig. 2d and e. The FWHM duration of the retrieved pulse shape is equal to 6.4 fs, which corresponds to approximately 2.4 cycles at the center wavelength (787 nm) – about twice as long as what has been used so far [11-14].

We simulate our results by a model suggested by Khurgin [20]. According to this model, the electric field of the pulse excites virtual carriers. The carrier generation is slightly different for positive and negative crystal momenta, owing to interference between different multiphoton processes leading to different parity states. This imbalance results in a charge $Q(t)$ transferred by the light electric field, which is proportional to the sum of the contributions from these multiphoton processes. These contributions can be expressed as integrals of an odd power of the vector potential $A(t)$, corresponding to the interference between excitation processes involving an odd and even number of photons. The total residual charge remaining after the pulse can be calculated as [20]

$$Q = \varepsilon_0 \chi^{(3)} \omega_0^4 \left( \int_{-\infty}^{\infty} d\tau\, A^3(\tau) + \frac{\chi^{(3)}}{\chi^{(1)}} \omega_0^2 \int_{-\infty}^{\infty} d\tau\, A^5(\tau) + \left[\frac{\chi^{(3)}}{\chi^{(1)}} \omega_0^2\right]^2 \int_{-\infty}^{\infty} d\tau\, A^7(\tau) + \ldots \right) S_{\text{eff}}, \quad (1)$$

where $\varepsilon_0$ is the vacuum permittivity, $\chi^{(1)}$ and $\chi^{(3)}$ denote the linear and third-order susceptibilities, $\omega_0$ is the central angular frequency of the laser pulse and $S_{\text{eff}}$ is the effective area of the junction. The electronic measurement circuit has a much slower response than the actual lightwave-driven currents inside the semiconductor. Hence, only the cumulative effect of the entire waveform can be detected and the integral in Eq. (1) spans over the complete simulated time window.

Results of this model are presented in Fig. 3 for two Gaussian pulses (red waveforms), centered at 800 nm, with a duration of 3.8 fs (1.4 cycles, Fig. 3a) and 6.5 fs (2.4 cycles, Fig. 3b). The sub-cycle charge (blue curve) follows the electric field shape of the laser pulse. For the short pulse (Fig. 3a), the waveform is rather asymmetric, which leads to a large charge transfer, remaining visible even after the pulse. It is this offset (marked by arrows) that is picked up by the external circuit. For the longer pulse duration of 6.5 fs, the degree of asymmetry is much smaller and the transferred charge at the end of the pulse approaches zero (Fig. 3b). However, a slight imbalance remains (see inset), which can still be detected by sensitive electronics. With a high repetition-rate laser system, a small charge transfer per pulse results in a rather large average current, in our case, in the nanoampere range (see Fig. 2a).

The laser pulse shown in Fig. 2d drives a charge transfer with a maximal amplitude of around 2 fC per laser shot (blue curve, Fig. 3c). The shaded area depicts the standard deviation from averaging multiple scans. We use the extracted pulse shape from the d-scan as the driving waveform in Eq. (1). Dispersion is added numerically at the center wavelength of the pulse. Using the experimentally determined pulse duration, focal spot size, optical power, and Fresnel coefficients for transmission, we estimate the driving field strength to 0.62 V/Å inside the material. The same peak field is used in the simulation, resulting in a very similar charge trace as measured experimentally (black curve, Fig. 3c).

We investigate the influence of the pulse duration experimentally by extending the parametric amplification process in our laser system to longer wavelengths, thus allowing us to obtain a shorter pulse with a duration of 5.7 fs (FWHM) measured by the d-scan technique. This pulse features only two optical cycles and results in a larger charge amplitude (blue curve, Fig. 3d). The modified pulse shape also leads to a more complex charge trace. We account for the shifted central wavelength and the modified pulse shape in the simulations and reproduce the measured signal in amplitude and shape (black curve, Fig. 3d). The only slight discrepancy is a phase jump in the simulation for large glass insertions, which we do not observe experimentally. This feature likely stems from the uncertainty of the exact spectral phase of the driving waveform in the model.

To investigate the nonlinearity of the generation process, we vary the electric peak field strength of our laser pulses. The lowest electric field at which we still detect appreciable currents above the noise floor is roughly 0.4 V/Å inside the sample. Increasing the field strength up to 0.6 V/Å increases also the charge amplitude, without affecting much the shape of the charge trace (Fig. 4a). We again model this experiment using the pulse shape extracted from a d-scan measurement and reproduce well the variation of the charge signal with field strength (Fig. 4b and c). The good agreement corroborates the multiphoton character of carrier excitation in our experiment. The data point at the highest field strength deviates from the expected scaling, which we attribute to accumulated damage. Prolonged exposure at these high intensity levels eventually leads to sample degradation, which is consistent with previous work [14].

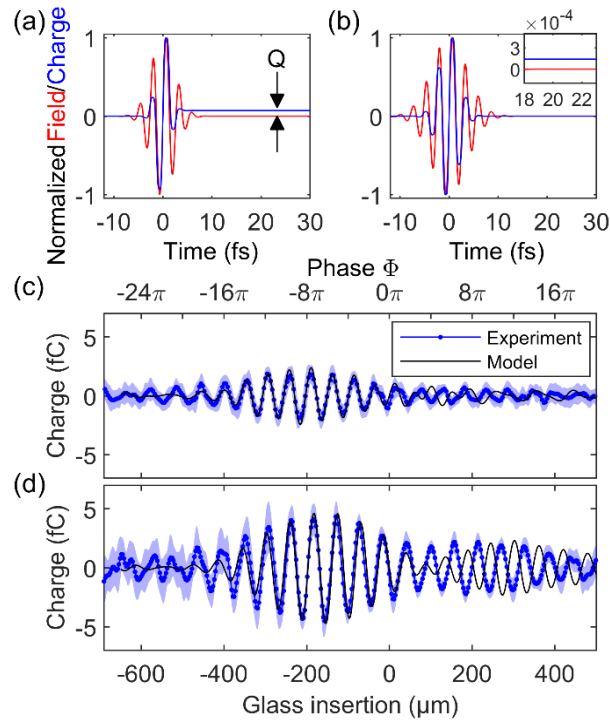

**Fig. 3. Influence of pulse duration.** (a) A model Gaussian pulse (red waveform) with 3.8 fs FWHM duration (1.4 cycles) drives a sub-cycle charge transfer (blue) in a semiconductor due to its asymmetry. A charge remains after the laser pulse (Q, marked by arrows), which can be measured by an external circuit. (b) A longer pulse of 6.5 fs duration (2.4 cycles) has a significantly lower asymmetry and drives a smaller, but still measureable current (see inset). (c) Measured charge signal (blue) as a function of glass insertion, employing the 6.4-fs pulses shown in Fig. 2. The shaded area depicts the standard deviation originating from averaging 20 traces. The black curve shows a model calculation utilizing the retrieved pulse profile from the d-scan. (d) Similar as (c) but for a shorter pulse duration of 5.7 fs.

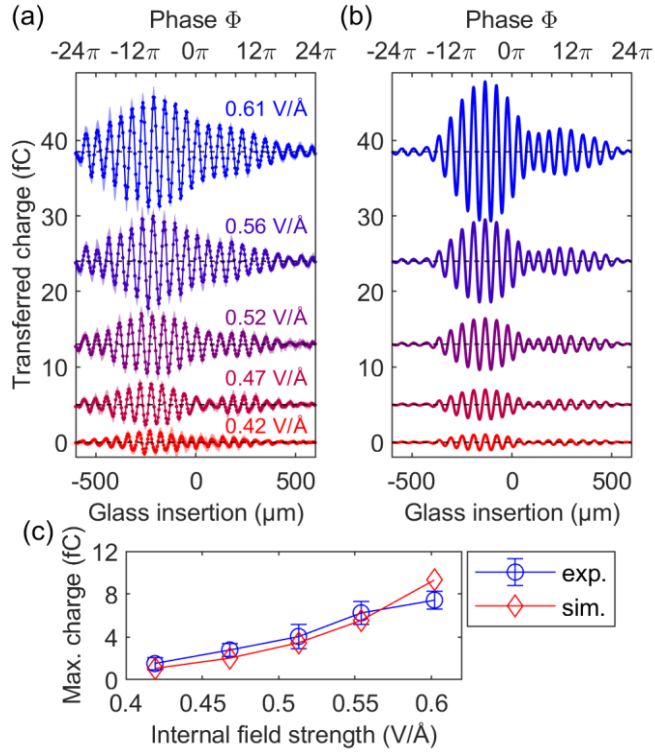

**Fig. 4. Electric field dependence of the transferred charge.** (a) Transferred charge per laser shot as a function of glass insertion or relative phase, respectively (Individual curves are offset). The data points are averaged over ten consecutive measurements, while the shaded area marks the standard deviation. (b) Simulated charge traces according to the model proposed by Khurgin, taking into account the experimental pulse shape. (c) Measured and simulated maximal charge signal as a function of electric field strength. Error bars depict the standard deviation of the measured charge.

## Conclusion

In conclusion, we have extended lightwave-driven currents to higher repetition-rate laser systems, while also doubling the duration of the used pulses. For accurate modelling of the transferred charge, a precise knowledge of the laser pulse shape is important. Conversely, lightwave-driven currents could be used to extract temporal properties of the driving pulses, without the need of a two-pulse pump-probe experiment [11,12,14]. The measured features can be well reproduced by a model solely based on interference of nonlinear effects of different parity, thus highlighting similarities to coherent control schemes of photo-injecting carriers [22-30]. Due to its wide use in conventional integrated electronics, GaN is an interesting candidate for implementing nano-structured devices or more complex electrode geometries, which potentially allow for lightwave-based logic.

**Funding.** The authors acknowledge funding from the Swedish Research Council, the European Research Council, Laserlab-Europe EU-H2020 654148 and the Knut and Alice Wallenberg Foundation. JV acknowledges funding from Marie Skłodowska-Curie Grant Agreement 793604 ATTOPIE.

**Disclosures.** JA: TEM Messtechnik GmbH (E). The remaining authors declare no conflicts of interest.